\documentclass[epjST]{svjour}
\usepackage{graphics}
\usepackage{amsmath}
\usepackage{color}
\usepackage{graphicx}

\usepackage{siunitx}
\usepackage{subfig}
\usepackage{array} 
\usepackage{amsmath,amssymb}

\begin{document}
	\newcommand{\sio}{{\mathrm{SiO}}}
	\newcommand{\sioo}{{\mathrm{SiO}_2}}
	\newcommand{\ion}{{\mathrm{ion}}}
	\newcommand{\bulk}{2}
	\newcommand{\etch}{{\mathrm{etch}}}
	\newcommand{\xa}{{x_\mathrm{a}}}
	\newcommand{\xb}{{x_\mathrm{b}}}
	\newcommand{\nox}{{n_\mathrm{ox}}}
	\title{Electro-Oxidation of p-Silicon in Fluoride-Containing Electrolyte: A Physical Model for the Regime of Negative Differential Resistance}
	\author{M. Salman\inst{1}\fnmsep\thanks{\email{munir.salman@tum.de}} \and M. Patzauer\inst{1} \and D. Koster \inst{2} \and 	F. La Mantia \inst{2} \and  K. Krischer \inst{1}\fnmsep\thanks{\email{krischer@tum.de}}}
	\institute{Technische Universit\"at M\"unchen, Physik Department, James-Franck-Stra\ss e 1, D-85748 Garching, Germany\and Universit\"at Bremen, Fachgebiet Energiespeicher- und Energiewandlersysteme, Bibliothekstrasse 1, D-28359 Bremen, Germany}
	\abstract{
		When Si is anodically oxidized in a fluoride containing electrolyte, an oxide layer is grown.
		Simultaneously, the layer is etched by the fluoride containing electrolyte.
		The resulting stationary state exhibits a negative slope of the current-voltage characteristics in a certain range of applied voltage.
		We propose a physical model that reproduces this negative slope.
		In particular, our model assumes that the oxide layer consists of both partially and fully oxidized Si 
		and that the etch rate depends on the effective degree of oxidation.
		Finally, we show that our simulations are in good agreement with measurements of the current-voltage characteristics, the oxide layer thickness, the dissolution valence, and the impedance spectra of the electrochemical system.
	} 
	\maketitle
	%
	\section{Introduction}
	
	When a silicon electrode is electrochemically dissolved in a fluoride containing electrolyte, it is covered by a layer of silica which can change its thickness in peculiar spatio-temporal patterns \cite{chazalviel1992ionic,lewerenz1993origin,lehmann1996origin,foll2009self,schonleber2012high,miethe2009irregular,miethe2012ellipsomicroscopic,proost2014origin}.
	(For a comprehensive overview of the work until 2003 see Chapter 5 in \cite{zhang2001electrochemistry}.)
	These spatio-temporal patterns make silicon in hydrofluoric solution one of the most relevant electrochemical systems for basic studies of non-linear dynamics.
	The most prominent example is the chimera state in which one part of the ensemble oscillates in synchrony, while the rest is turbulent \cite{abrams2004chimera,schmidt2014coexistence,schonleber2014pattern,schmidt2015chimeras,patzauer2017autonomous}.
	Although it is possible to find some attempt of modeling \cite{schonleber2016comparison,zensen2014capacitance,chazalviel1992theory}, not even the simplest, periodic, spatially homogeneous oscillations have so far been consistently and convincingly ascribed to a concrete physical mechanism.
	
\begin{figure}
		\centering
		\includegraphics[width=0.5\textwidth]{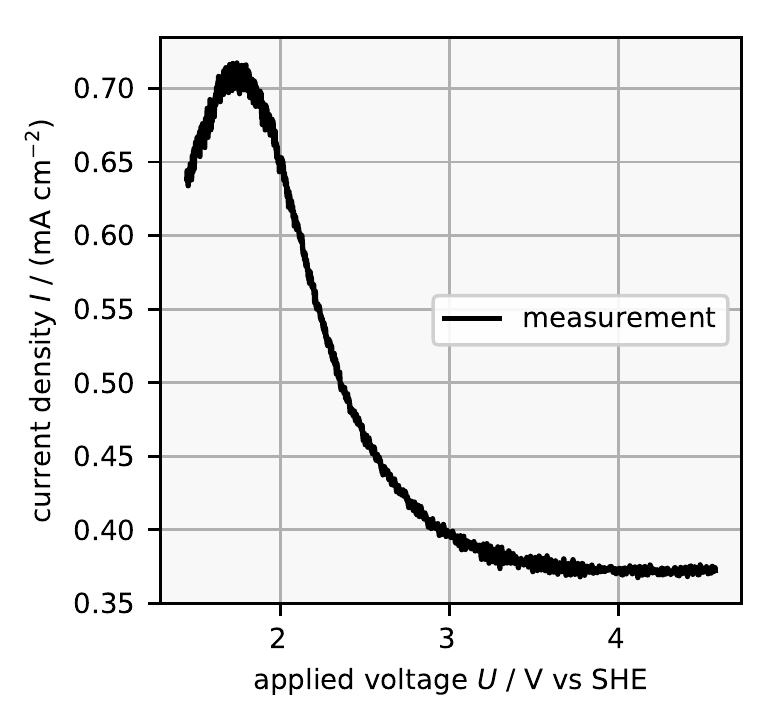}
		\caption{1\,mV/s cyclic voltammogram of a (111) p-silicon electrode of 5 to 25\,$\mathrm{\Omega}\,\text{cm}$ in a solution of 0.05\,M NH$_4$F and 0.025\,M H$_2$SO$_4$ (pH 2.3) }
		\label{fig:introCV}
	\end{figure}

	When trying to model the system one has to consider the following processes. 
	Applying a postive voltage to a silicon electrode in an aqueous electrolyte results in the formation of silicon dioxide \cite{zhang2001electrochemistry}. 
	As the oxidation reaction proceeds, the resulting layer of oxide eventually passivates the electrode, inhibiting further reaction. 
	However, if the solution contains HF or HF$_2^-$ species, the oxide layer is etched away \cite{uhlir1956electrolytic,osseo1996dissolution}. 
	The oxide layer thickness may, thus, attain a steady state at which the rates by which the oxide layer grows into the silicon electrode and is dissolved by the fluoride containing solution balance exactly.
	The current voltage characteristics of this equilibrium is very similar for a wide range of parameters \cite{chazalviel1991voltammetric} and can exemplarily be seen in Figure~\ref{fig:introCV} for voltages positive to the porous oxide formation. 
	The peak marks the voltage above which an oxide layer is formed \cite{eddowes1990anodic,blackwood1992electrochemical}.
	This equilibrium can become unstable at certain experimental parameter values which then leads to spatio-temporal variations of the oxide layer thickness.
	
	In this article we follow the mindset that complicated dynamics bifurcate from simple dynamics, and hence that the first step towards a comprehensive understanding must be to find out which physical processes lead to the equilibrium properties, for example the stationary current voltage characteristics and the electrochemical impedance.
	So far there is a model for the current voltage characteristics in the electropolishing regime at voltages below the current peak in Figure~\ref{fig:introCV} \cite{cheggou2009theoretical}. 
	There is also a more abstract model for the electrochemical impedance in the resonant voltage regime \cite{chazalviel1992theory}.
	We present a physical model that explains both the current voltage characteristics and the impedance spectrum in the regime of negative differential resistance.
	Moreover, it reproduces the dependence of the oxide layer thickness and the dissolution valence on the applied voltage.
	
	The central idea of our model is that the oxide is etched with a rate that depends on its composition. 
	This assumption was motivated by measurements of dissolution valence \cite{eddowes1990anodic,blackwood1992electrochemical,schonleber2012high}, measurements of the current voltage characteristics, and by the fact that the etching is likely to occur purely chemically, i.e. without charge transfer \cite{memming1966anodic,eddowes1990anodic,cattarin2000electrodissolution}.
	Then, na\"ively, the applied voltage should not influence the etching of the oxide layer.
	However, the rates of the purely chemical etching and the voltage dependent oxidation have to be the same at the steady state.
	Hence, there has to be an indirect dependence of the etch rate on the applied voltage.
	We suggest that this indirect dependence can be explained by the change in oxide composition as the voltage is varied.
	This change could arise due to the accumulation/depletion of ions or defects moving through the oxide layer.
	In this article we propose a physical model incorporating this idea.
	
	First we explain our applied methods in Section~\ref{sec:methods}. The model is described and defined in Section~\ref{sec:model}.
	In Section~\ref{sec:predictions} the predictions of the model are compared to the experimental data which we obtained from cyclic voltammmetry, ellipsometry, and dynamic multi-frequency analysis as well as to values of the dissolution valence from the literature \cite{eddowes1990anodic,blackwood1992electrochemical,schonleber2012high}. Finally, we conclude the paper in Section~\ref{sec:conclusion}.

	\section{Methods}
	\label{sec:methods}
	
	\subsection{Experimental methods}
	
	All electrochemical measurements were performed with p-doped silicon electrodes ((111) orientation, 5-25\,$\mathrm{\Omega}\,\text{cm}$). Ohmic back contact preparation, surface pretreatment and cleaning were done as described in \cite{patzauer2017autonomous}. The aqueous electrolyte was prepared from ultra pure water (ELGA Purelab Ultra, 18.2$\,$M$\mathrm{\Omega}\,$cm, TOC $<\,$3ppb), NH$_4$F (Merck, p.a.) and H$_2$SO$_4$ (Merck, Suprapur). For all experiments shown here, the electrolyte contained 0.05~M NH$_4$F and the pH was adjusted to 2.3, according to dissociation constants from the literature \cite{cattarin2000electrodissolution}. The counter electrode was a Pt wire and a sat. Hg/Hg$_2$SO$_4$ electrode served as reference elecrode. 
	
	The in-situ ellipsometric measurements in Figure~\ref{fig:emsi} and the cyclic voltammogram in Figure~\ref{fig:introCV} and \ref{fig:cv} were obtained with the spectroelectrochemical set-up described in \cite{miethe2012ellipsomicroscopic}. In short, the stationary working electrode is illuminated with elliptically polarized, blue light from an LED with 470\,nm wavelength through a window in the cell wall.
	After reflection at an incident angle of approximately 70\,$^\circ$ the polarization of the light has changed, depending on the optical path through the oxide.
	The change in polarization is converted to an intensity signal by letting the light pass through a polarizer and is measured using a CCD camera.
	The experiments were controlled with an FHI potentiostat (Fritz-Haber-Institut, Elab), and the electrolyte was stirred with a magnetic stirrer.
	
	The dynamic impedance measurements \cite{battistel2016analysis,koster2017dynamic} in Figures~\ref{fig:exp_nyq},~\ref{fig:exp_bode},~\ref{fig:NyqSimVsExp} were done using a rotating disk electrode 
	set up (AFMSRCE, Pine Research Instrument, Inc) with a disc diameter of 5 mm.
	A multisine perturbation was generated with a 2-channel waveform generator (Keysight) and added to the potential applied by the potentiostat (BioLogic Science Instruments, SP-300) to the working electrode.  For the evaluation of the data we employed dynamic multi-frequency analysis as outlined in \cite{battistel2016analysis,koster2017dynamic}. Further experimental details of the impedance measurements and the cell used can be found in \cite{koster2018measurement}.
	
	\subsection{Numerical methods}
	
	The numerical results were obtained with the proprietary finite element software COMSOL 5.2 \cite{comsol2016comsol}. 
	We solved partial differential equations numerically on a one dimensional spatial domain perpendicularly to the electrode, which represents the oxide layer of finite thickness.
	From the available COMSOL-physics-components we used the ``General Form PDE" component for chemical species, the ``Poisson's Equation" component for the electrostatic potential, and ``Moving Mesh" component to simulate the growth and dissolution of the oxide layer. The spatial domain was split into 2000 finite elements of the same length, which were adapted continuously as the boundaries moved.
	All simulations were run with the COMSOL-study ``Time Dependent" and all solver configurations were left at their default except for the ``Relative tolerance" which we reduced to $10^{-7}$ for the electric impedance calculations and to $10^{-5}$ in all other cases.
	All the parameters, variables and physical constants used in the simulations are listed in Tables~\ref{tab:param}, \ref{tab:var}, \ref{tab:const}.
	\begin{table}
		\centering
		$\begin{array}{c|c|l}
		\mathrm{Notation}&\mathrm{Value}&\mathrm{Meaning}\\
		\hline
		\nox 	& 4.381\cdot 10^{-2}\,\text{mol\,cm}^{-3} 	&\text{molar oxide density} \\
		n_\ion^0 & 9.36\cdot 10^{-5}\,\text{mol\,cm}^{-3} 	&\text{O$^{2-}$ concentr. in equilib. with solution}  \\
		D_\ion 	& 1.1962\cdot 10^{-13}\,\text{cm}^2\,\text{s}^{-1} 	&\text{diffusion coefficient of O$^{2-}$} \\
		D_\sio 	& 3.2\cdot 10^{-14}\,\text{cm}^2\,\text{s}^{-1}	&\text{diffusion coefficient of SiO } \\
		k_1 	& 1600\,\text{cm\,s}^{-1} 	&\text{reac. coeff. of first oxidation step}\\
		k_\bulk & 2.048\cdot10^{4}\,\text{cm}^3\text{mol}^{-1}\,\text{s}^{-1}	&\text{reac. coeff. of second step}\\
		k_3 	& 3\cdot10^{15}\,\text{cm}^4\text{mol}^{-1}\text{s}^{-1} 	&\text{reac. coeff. of direct SiO}_2\text{ production}\\
		U 	& 2\,\text{V}	&\text{electrode voltage with offset} \\
		\varepsilon_\mathrm{ox} & 1\cdot\varepsilon_0 &\text{permittivity of mixed oxide} \\
		\eta_\sio & 6\cdot \eta_\sioo	&\text{etch velocity of pure SiO}	\\	
		\eta_\sioo & 2.09\cdot 10^{-8}\,\text{cm\,s}^{-1}	&\text{etch velocity of pure SiO}_2	\\
		
		\end{array}$
		\caption{Default parameters used if not stated otherwise}
		\label{tab:param}	
	\end{table}
	\begin{table}
		\centering
		$\begin{array}{c|c|l}
		\mathrm{Notation}&\mathrm{Unit}&\mathrm{Meaning}\\
		\hline
		I(t)	& [\text{A}\text{\,m}^{-2}] & \text{externally measurable electric current}\\
		I_\mathrm{ch}(t)	& [\text{A}\text{\,m}^{-2}] & \text{current connected to capacitive charging}\\
		I_\mathrm{reac}(t)	& [\text{A}\text{\,m}^{-2}] & \text{current connected to reactions}\\
		J_\ion(x,t)& [\text{mol\,m}^{-2}\text{\,s}^{-1}]	&	\text{flux of O$^{2-}$}\\
		J_\sio(x,t)& [\text{mol\,m}^{-2}\text{\,s}^{-1}]&	\text{flux of SiO}\\
		n_\ion(x,t)& [\text{mol\,m}^{-3}]&	\text{concentration of O$^{2-}$}\\
		n_\sio(x,t)& [\text{mol\,m}^{-3}] &	\text{concentration of SiO}\\
		q(t)&[\text{C\,m}^{-2}]&\text{Si space charge plus surface charge}\\
		r_1(t)	& [\text{mol\,m}^{-2}\text{\,s}^{-1}] &	\text{rate of the reaction in Eq.~\eqref{reac:1}}\\
		r_2(x,t)& [\text{mol\,m}^{-3}\text{\,s}^{-1}] &	\text{rate density of the reaction in Eq.~\eqref{reac:2}}\\
		r_3(t)	& [\text{mol\,m}^{-2}\text{\,s}^{-1}] &	\text{rate of the reaction in Eq.~\eqref{reac:3}}\\
		r_\etch(t)& [\text{mol\,m}^{-2}\text{\,s}^{-1}] &	\text{rate of etching} \\
		t	& [\text{s}] &	\text{time coordinate}\\
		x	& [\text{m}] &	\text{spatial coordinate perpendicular to surface}\\
		\xa(t)	& [\text{m}] &	\text{position of the Si/SiO$_x$ interface}\\
		\xb(t)	& [\text{m}] &	\text{position of the SiO$_x$/solution interface}\\
		\nu(t) & [1] & \text{dissolution valence, i.e. e}^-\text{ transferred per Si dissolved}\\
		\varphi(x,t)& [V] &	\text{electrostatic potential}\\
		\Delta\varphi_\mathrm{WE}(t) & [V]	& \text{voltage between working and reference electrode}\\
		\end{array}$
		\caption{Variables}
		\label{tab:var}
	\end{table}
	\begin{table}
		\centering
		$\begin{array}{c|c|l}
		\mathrm{Notation}&\mathrm{Value}&\mathrm{Meaning}\\
		\hline
		F	& 96.485\,\text{C mol}^{-1}	&\text{Faraday constant}\\
		T	&	293\,\text{K}	& \text{assumed room temperature}\\
		z_\ion	& -2 & \text{elementary charges per O$^{2-}$}\\
		\varepsilon_0 & 8.854\cdot 10^{-12}\,\text{F\,m}^{-1}&\text{vacuum permittivity}\\
		\end{array}$
		\caption{Physical constants}
		\label{tab:const}
	\end{table}

	
	\section{Model description}
	\label{sec:model}
	
	In this Section we model the oxide layer, SiO$_\mathrm{x}$, that covers the Si anode in hydrofluoric solution. 
	We aim to reproduce the negative differential resistance, i.e. $\mathrm{d}I/\mathrm{d}\varphi_\text{WE}<0$, where $I$ is the electric current and $\varphi_\text{WE}$ is the potential of the working electrode.
	
	\subsection{Concept}
	\label{sec:concept}
	
\begin{figure}
		\centering
		\includegraphics[width=\textwidth]{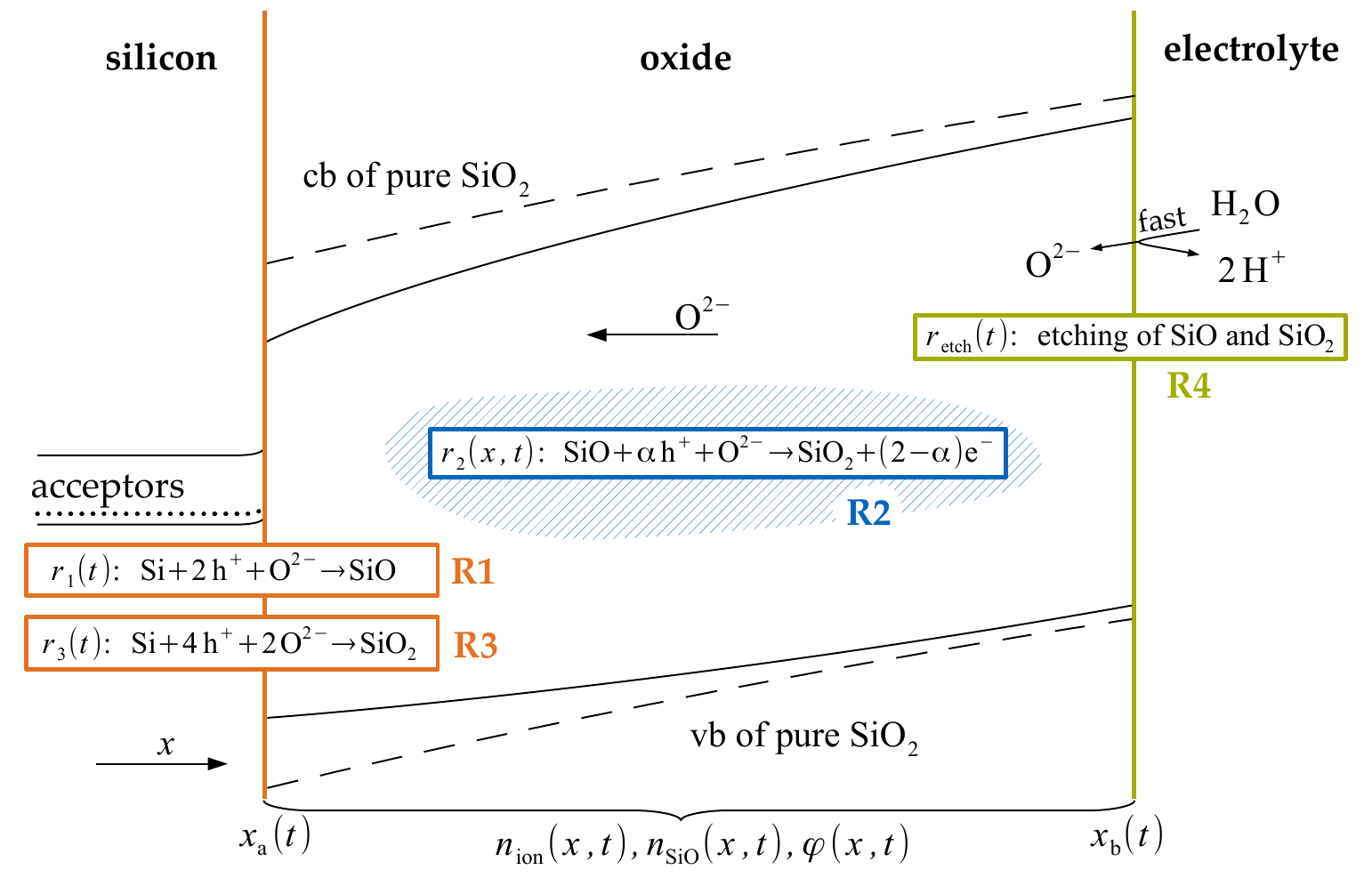}	
		\caption{A sketched band diagram of the model described in Section~\ref{sec:model} with all the variables that define a state: Interface positions $\xa(t)$ and $\xb(t)$, concentration of O$^{2-}$ ions $n_\ion (x,t)$, concentration of intermediate oxide $n_\sio (x,t)$ and electric potential $\varphi(x,t)$. The bandgap of SiO$_x$ (solid black line) is smaller than that of pure SiO$_2$ (dashed black line) \cite{lin2004suboxide}. 
			Furthermore, the reaction that allows O$^{2-}$ to enter the oxide layer is assumed to be fast compared to the transport inside the layer, so that the chemical potential of O$^{2-}$ at $\xb(t)$ is prescribed by the electrolyte.}
		\label{fig:sketch}
	\end{figure}

	We assume the electrodissolution of Si to occur in three steps which are sketched in Figure~\ref{fig:sketch}. 
	The initial step is silicon oxidation at the interface between the silicon and the oxide layer. 
	In this step, the oxide layer grows into the silicon electrode. 
	The pure silicon is, thus, replaced by a partially oxidized silicon species, e.g. SiO. 
	
	The second step happens inside the layer volume and transforms the partially oxidized silicon, SiO, into the final oxidation state, SiO$_2$. 
	The transformation from Si to SiO$_2$ is likely to involve several sub-steps and further intermediate oxidation states, correspondingly. 
	Our model merges these sub-steps for simplicity.
	In the third and final step the layer is etched away purely chemically.
	The etch rate is assumed to depend on the oxide composition and to be faster if the fraction of partial oxide is higher. 
	
	One possible way to realize the mechanism we stated above is given by the following set of reactions. Ions O$^{2-}$ enter the oxide layer by a reaction like:
	\begin{align}
		(\mathrm{H}_2\mathrm{O})_\text{solution} ~& \leftrightharpoons ~~~ 2(\mathrm{H}^+)_\text{solution} + (\mathrm{O}^{2-})_\text{layer} 
		\label{eq:equilibriumreaction}
		\tag{R0}
	\end{align}
	Under the present high electric fields (V/nm) the ions travel through the layer and oxidize the silicon in two steps
	\begin{align}
		\mathrm{Si+2h}^+ + \mathrm{O}^{2-}   ~&\stackrel{k_1}{\rightarrow}~~~ \mathrm{SiO} \label{reac:1} \tag{R1}\\ 
		\mathrm{SiO}+\alpha \mathrm{h}^+ + \mathrm{O}^{2-}  ~&\stackrel{k_2}{\rightarrow}~~~ \mathrm{SiO}_2 + (2-\alpha)\mathrm{e}^-,
		~~~~\text{with }0<\alpha<2
		\label{reac:2}
		\tag{R2}
	\end{align}
	or in just a single step
	\begin{align}
		\mathrm{Si+4h}^+ + 2 \mathrm{O}^{2-}   ~&\stackrel{k_3}{\rightarrow}~~~ \mathrm{SiO}_2 \label{reac:3}  
		\tag{R3}
	\end{align}
	Reaction \eqref{reac:1}, occurs at rate $r_1(t)$ at the boundary between silicon and a mixed layer of SiO and SiO$_2$. 
	This mixture is referred to as SiO$_x$ in the following.
	Inside the mixed oxide layer the SiO created in Reaction~\eqref{reac:1} can be further oxidized to SiO$_2$ in Reaction~\eqref{reac:2} at rate $r_2(x,t)$.  
	Alternatively, Si can be directly oxidized to SiO$_2$ in Reaction~\eqref{reac:3} at rate $r_3(t)$.
	
	The last step is the etching of oxide at the boundary between oxide and solution, named SiO$_x$/solution interface, with a rate $r_\etch(t)$ depending on the oxide composition right at the boundary. The corresponding reaction reads
	\begin{align}
		\mathrm{SiO}_x + 6 \mathrm{HF} ~& \rightarrow ~~~ \mathrm{SiF}_6^{2-} + x\, \mathrm{H}_2\mathrm{O} + 2 \mathrm{H}^+ + (2-x)\mathrm{H}_2
		\label{eq:dissolution}
		\tag{R4}
	\end{align}
	with some $x$ between 1 and 2. The dissolution of SiO$_2$ in hydrofluoric solution is well investigated \cite{osseo1996dissolution,cattarin2000electrodissolution} and more complicated than Reaction~\eqref{eq:dissolution}, but our simplification will serve the purpose. 
	
	Although the reactions that we assumed above lead to good predictions, the mechanism could probably also be implemented with other reactions.
	One could for example consider OH$^-$ ions and Si(OH)$_x$ instead of O$^{2-}$ ions and SiO$_x$.
	
	\subsection{Kinetic description}
	\label{sec:kinetics}
	We transcribed the ideas from Section \ref{sec:concept}, which are sketched in Figure~\ref{fig:sketch}, to a set of coupled partial differential equations of time $t$ and of one spatial dimension $x$ which points perpendicularly away from the electrode surface.
	These partial differential equations describe the transport and the reactions inside the oxide layer and at its boundaries.
	As etching \eqref{eq:dissolution}, reaction \eqref{reac:1}, and reaction \eqref{reac:3} proceed, the corresponding interfaces move. 
	Let $\xa(t)$ and $\xb(t)$ be the time-dependent position of the Si/SiO$_x$ and SiO$_x$/solution interfaces, respectively. 
	Assuming a constant molar oxide density $\nox$ for simplicity, the velocities can be written in terms of the respective reaction rates:
	\begin{align}
		\partial_t \xa(t) ~&=~~ -\frac{1}{\nox} (r_1(t)+r_3(t)) \label{eq:xa}\\
		\partial_t \xb(t) ~&=~~ -\frac{1}{\nox} r_\etch(t) \label{eq:xb}
	\end{align}
	where the rates per unit area $r_1(t)$, $r_3(t)$ and $r_\etch(t)$ are defined below in Equation \eqref{eq:reac1}, \eqref{eq:reac3} and \eqref{eq:reac2}.
	
	Since we exclusively consider situations where the silicon surface is covered with some (though possibly thin) oxide layer, the rates $r_1(t)$ and $r_3(t)$ are assumed to be limited solely by a lack of O$^{2-}$ ions. 
	Silicon and h$^+$, the other reactants involved, are present in high concentrations because we consider p-type silicon under anodic bias.
	The etch rate $r_\etch(t)$ for a certain oxide mixture is linearly interpolated between the etch rate of pure SiO$_2$  and the etch rate of pure SiO with proportionality coefficients $\eta_\sioo$, $\eta_\sio$:
	\begin{align}
		r_1(t) 	~&=~~ k_1 n_\ion(\xa(t),t)\label{eq:reac1}\\
		r_3(t) 	~&=~~ k_3 n_\ion(\xa(t),t)^2\label{eq:reac3}\\
		r_\etch(t) 	~&=~~ \eta_\sio n_\sio(\xb(t),t)+\eta_\sioo \left[\nox-n_\sio(\xb(t),t)\right]\label{eq:reac2}
	\end{align}
	where $\nox$ is the molar density of silicon in the oxide, and $n_\sio(x,t)$, $n_\ion(x,t)$ are the molar densities of SiO and O$^{2-}$ respectively. The concentration of SiO$_2$ is expressed as $\nox-n_\sio(x,t)$.
	Hereby we assume SiO$_2$ to be the non-SiO fraction of the entire molar oxide density $\nox$. 
	The rate coefficients $k_1$ and $k_3$ depend on the potential drop across the space charge layer which is assumed to be constant in our model, because the differential capacity of the space charge layer is large under forward bias \cite{sze2006physics}. 
	Thus, $k_1$ and $k_3$ are constant as well.
	
	Unlike the reactions \eqref{reac:1} and \eqref{reac:3} which are interface reactions, reaction \eqref{reac:2} occurs inside the oxide volume. 
	From the participating species we expect electrons/holes to leave/enter the oxide quickly via direct tunneling or trap hopping which has been observed in the electroluminescence spectra of electrically biased oxide films \cite{hasegawa1988anodic}. 
	The remaining degrees of freedom in the oxide layer are then the concentrations of SiO and O$^{2-}$. 
	The dynamics of these two species is dictated by their respective mass balance equations which contain transport and reaction terms:
	\begin{align}
		\partial_t n_\ion (x,t) ~&=~~ -\nabla J_\ion (x,t) - r_\bulk(x,t) \label{eq.ion} \\
		\partial_t n_\sio (x,t) ~&=~~ -\nabla J_\sio (x,t) - r_\bulk(x,t)    \label{eq.sio}
	\end{align}
	where $J_\ion (x,t)$ and $J_\sio (x,t)$ are the molar fluxes of O$^{2-}$ and SiO respectively. 
	
	Let $\varphi(x,t)$ be the electrostatic potential. 
	Since O$^{2-}$ is the only charged species considered inside the oxide, the total charge density in the layer is $z_\ion F n_\ion(x,t)$, where $z_\ion$ is the number elementary charges per ion, i.e. -2, and $F$ is the Faraday constant. Poisson's equation thus reads:
	\begin{equation}
		\nabla^2 \varphi(x,t) ~=~ -\frac{z_\ion F}{\varepsilon_\mathrm{ox}}n_\ion(x,t)
	\end{equation}
	Introducing the diffusion constants $D_\ion, D_\sio$ and the rate coefficient $k_2$, the transport and reaction terms in Equation \eqref{eq.ion} and \eqref{eq.sio} can be modeled as:
	\begin{align}
		J_\ion(x,t)	~&=~~ -D_\ion\nabla n_\ion(x,t) + \frac{D_\ion z_\ion F}{RT}n_\ion(x,t)\nabla \varphi(x,t) \label{eq:Jion}\\
		J_\sio(x,t)	~&=~~ -D_\sio\nabla n_\sio(x,t) \label{eq:Jsio}\\
		r_\bulk(x,t) ~&=~~ k_\bulk n_\ion(x,t) n_\sio(x,t) \label{eq:R2}
	\end{align}
	where diffusive transport was assumed to be Fickian in the Equations \eqref{eq:Jion},\eqref{eq:Jsio}  and the Einstein-Smoluchowski relation was used to derive the ionic migration term in Equation~\eqref{eq:Jion}. $R$ and $T$ are the gas constant and temperature.
	
	The diffusion of SiO reflects the thermal restructuring of the solid oxide which we expect to be relevant on the considered length scales of some nanometers, after trying various diffusion coefficients for SiO.
	
	\subsection{Boundary conditions}
	\label{sec:bc}
	Continuity yields Equation~\eqref{eq:boundLeftIon} and \eqref{eq:boundLeftSiO} as boundary conditions for the Si/SiO$_x$ interface at $\xa$:
	\begin{align}
		J_\ion(\xa(t),t) ~&=~~ -r_1(t)-2 r_3(t)+n_\ion(\xa(t),t)\,\partial_t\xa(t) \label{eq:boundLeftIon}\\
		J_\sio(\xa(t),t) ~&=~~~ r_1(t)+n_\sio(\xa(t),t)\,\partial_t\xa(t) \label{eq:boundLeftSiO}
	\end{align}
	For the SiO$_x$/solution interface at $\xb(t)$ we assume a certain fixed  concentration $n_\ion^0$ of O$^{2-}$ that is tantamount to assuming that Reaction \eqref{eq:equilibriumreaction} is always in equilibrium.
	The oxide dissolution is accounted for by the movement of the SiO$_x$/solution interface in Equation \eqref{eq:xb}. 
	Thus, the flux of SiO at the boundary should be set to zero:
	\begin{align}
		n_\ion(\xb(t),t) ~&=~~ n_\ion^0 \label{eq:boundRightIon}\\
		J_\sio(\xb(t),t) ~&=~~ 0 \label{eq:boundRightSiO}
	\end{align}
	
	To come up with the boundary conditions for the electrostatic potential $\phi(x,t)$ we made the following simplifying assumption about the space charge layer.
	The differential capacity of the space charge layer of p-type silicon under strong anodic bias is very large, like in the case of a metal-insulator-semiconductor capacitor \cite{sze2006physics}. 
	The same is true for the Helmholtz layer which also has a large capacity.
	Therefore, any change in the total applied voltage will drop on the much smaller capacitor formed by the oxide layer.
	Thus, the oxide sees the externally applied voltage with an offset that is fixed. 
	Under these assumptions we end up with the following boundary conditions for the electrostatic potential:
	\begin{align}
		\varphi(\xa(t),t) ~&=~~ U(t) \label{eq:phiLeft}\\
		\varphi(\xb(t),t) ~&=~~ 0 \label{eq:phiRight}
	\end{align}
	where $U(t)$ equals the externally applied voltage plus/minus a constant offset.
	
	\subsection{Total electric current}
	The total electric current $I(t)$ that flows into the working electrode is the sum of the reaction current $I_\mathrm{reac}(t)$ and capacitive charging current $I_\mathrm{ch}(t)$. 
	\begin{equation}
		I(t)=I_\mathrm{reac}(t)+I_\mathrm{ch}(t)
	\end{equation}
	The current $I(t)$ is used in Section~\ref{sec:predictions} to predict a cyclic voltammogram and an impedance spectrum and compare them to electrochemically measured data.
	
	The reaction current $I_\mathrm{reac}(t)$ is calculated from the reaction rates $r_1(t)$, $r_3(t)$ and $r_2(x,t)$, where $r_1(t)$,$r_3(t)$ were defined as rates per surface and $r_2(x,t)$ as a rate per volume.
	\begin{equation}
		I_\mathrm{reac}(t)=2F\left[r_1(t)+2\cdot r_3(t)+\int\limits_{\xa(t)}^{\xb(t)}\mathrm{d}x\,r_2(x,t)\right]
	\end{equation}
	Remember that we assumed that the electrons/holes created/consumed in the bulk at rate $r_2(x,t)$ leave/enter the oxide via trap hopping or direct tunneling, more or less immediately. 
	Thus, the current resulting from $r_2(x,t)$ is contributing to $I_\mathrm{reac}(t)$ without any further delay.
	
	The capacitive charging current $I_\mathrm{ch}(t)$ is calculated from the electric potential $\varphi(x,t)$ as
	\begin{equation}
		I_\mathrm{ch}(t) = 
		-\varepsilon_\mathrm{ox}\,\partial_t \left[\lim\limits_{h \searrow 0} \partial_x \varphi(\xa(t)+h,t)\right]
		\label{eq:current}
	\end{equation}  
	which is simply the time derivative of the integrated space charge in the semiconductor. Equation~\eqref{eq:current} was obtained by integrating Poisson's equation from $-\infty$ to $\xa(t)+h$ and differentiating both sides of the equation by $t$.

	\subsection{Discussion on free parameters}
	\label{sec:discussion}
	
	There are quite some tunable parameters in the model.
	The electrochemical potential of ions in equilibrium and the transport and reaction rates were chosen somewhat arbitrarily, but for the remaining parameters one can make an educated guess.
	For the molar density $n_\mathrm{ox}$ and the permittivity $\varepsilon_\mathrm{ox}$ one can approximately take known values of thermal oxide.
	The etch rates of partial and stoichiometric oxide $\eta_\sio,\,\eta_\mathrm{SiO_2}$ can be roughly estimated from the experimentally observed etch rates.
	The offset between the model voltage parameter $U$ and the actual voltage across the interface can be estimated considering the dependence of the space charge layer's differential capacity on the voltage drop across it.
	
	\section{Results and discussion}
	\label{sec:predictions}
	
	\subsection{Current-voltage characteristics}
	\label{sec:cv_character}
	The model is supposed to reproduce the negative slope in the current-voltage characteristics. 
	In an electrochemical set up, this negative slope is measured between the broad current peak and the resonant current plateau in Figure~\ref{fig:introCV}, if the applied voltage is cycled.
	We simulated such cyclic voltammetry by making the model parameter $U$ a periodic triangular function of time, since $U$ represents the potential difference between the silicon electrode and the electrolyte.
	The parameter $U$ differs from the actual applied voltage by an offset due to the space charge layer and the Helmholtz layer, which are discussed in Section~\ref{sec:bc}, but are not quantified in this model.
	
\begin{figure}
		\centering
		\includegraphics{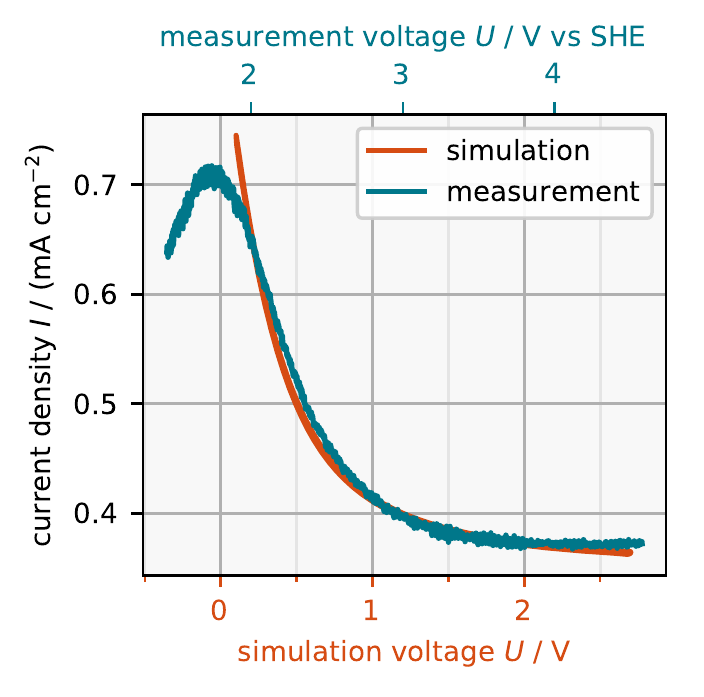}
		\caption{Comparison of simulated and measured cyclic voltammograms. Orange: Cyclic voltammogram obtained from the model described in Section~\ref{sec:model} at a scan rate of 1\,mV/s, using the parameters in Table~\ref{tab:param}. Blue: Cyclic voltammogram at  1\,mV/s of a (111) p-silicon electrode of 5 to 25\,$\mathrm{\Omega}\,\text{cm}$ in a solution of 0.05\,M NH$_4$F and 0.025\,M H$_2$SO$_4$ (pH 2.3) }
		\label{fig:cv}
	\end{figure}

	After an initial transient the resulting voltammogram is robust for a wide range of initial conditions.
	Figure~\ref{fig:cv} shows a simulated voltammogram with a scan rate of 1\,mV\,s$^{-1}$ and the corresponding electrochemical measurement.
	In both curves the current decreases with higher electrode potential.
	With faster scan rates we observed that the hysteresis of both the simulated and the electrochemically measured cyclic voltammogram become larger and that the measured clockwise direction of the hysteresis is matched in the simulations.
	If the scan rate is low enough, the simulated cyclic voltammogram converges to a single line, corresponding to the stationary current voltage curve, again in agreement with experiments.

	\subsection{Oxide layer properties}
\begin{figure}
		\centering
		\includegraphics{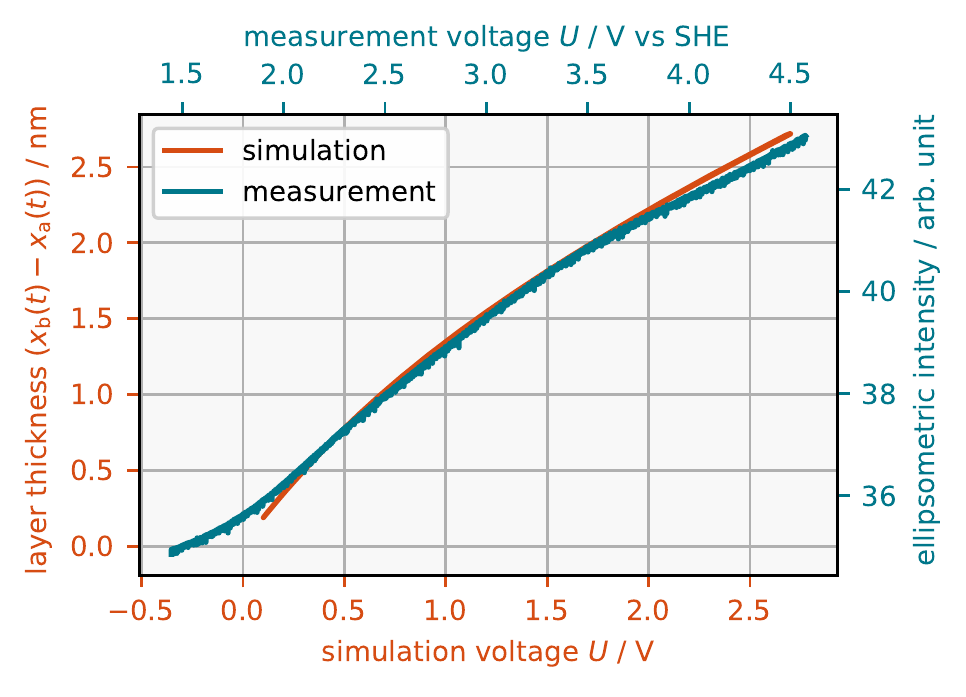}
		\caption{Comparison of the simulated oxide layer thickness during a slow voltage scan and the corresponding experimental in-situ ellipsometric signal which is proportional to the layer thickness \cite{miethe2012ellipsomicroscopic}. Orange: Oxide layer thickness vs. voltage $U$ during simulated cyclic voltammetry using the model described in Section\,\ref{sec:model} at a scan rate of 1\,mV/s, with the parameters of table\,\ref{tab:param}. Blue: Ellipsometric intensity representing the oxide layer thickness during a 1\,mV/s cyclic voltage scan of (111) p-silicon of 5 to 25\,$\mathrm{\Omega}\,\text{cm}$ in a solution of 0.05M NH$_4$F and 0.025M H$_2$SO$_4$ (pH2.3)}
		\label{fig:emsi}
	\end{figure}

	The thickness of the oxide layer as obtained during the simulated cyclic voltammetry is plotted in Figure~\ref{fig:emsi}, together with the corresponding electrochemical measurement which shows an in-situ ellipsometric signal.
	The ellipsometric signal is approximately proportional to the oxide layer thickness but also depends on the refractory index inside the layer and in front of it.
	It can be used as a qualitative estimation of the layer thickness \cite{miethe2012ellipsomicroscopic}.
	In both plots the thickness increases with the applied voltage.
	
\begin{figure}[htbp]
		\subfloat[Stationary concentration of O$^{2-}$ ions $n_\ion(x,t)$ (dashed line) and of partially oxidized silicon $n_\sio(x,t)$ (solid line) for different applied voltages.]{\includegraphics{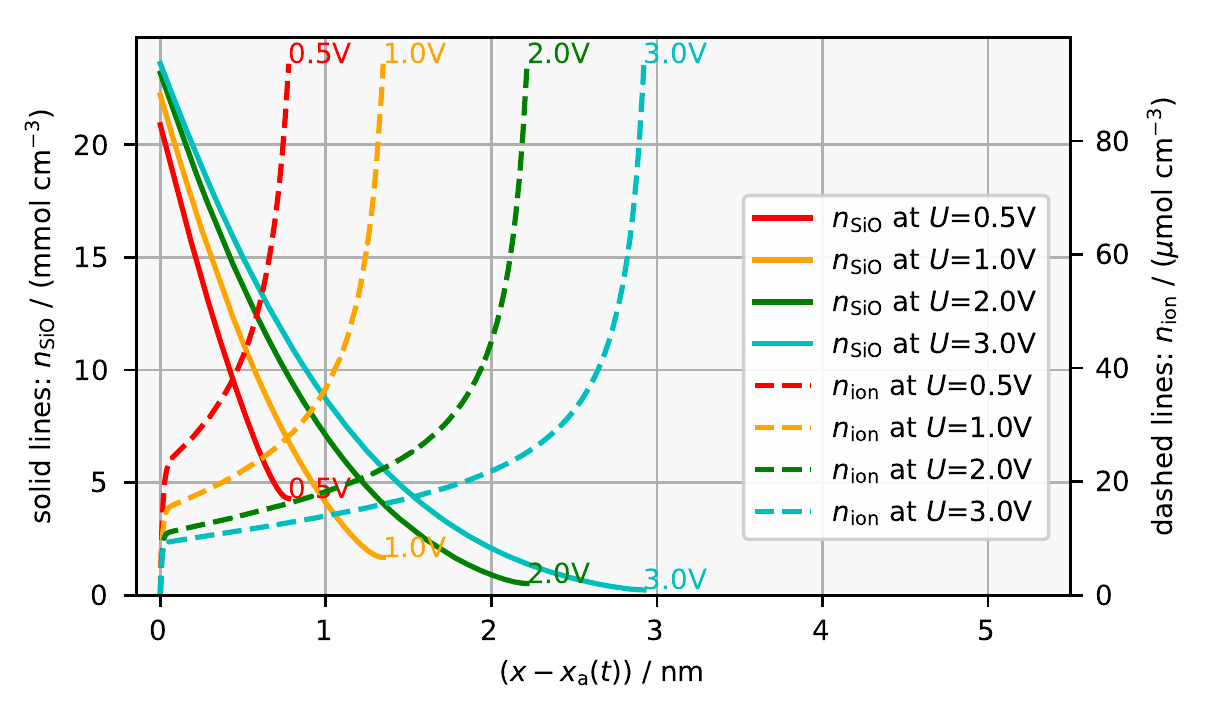}\label{fig:profiles_n}}\\
		\subfloat[Stationary profile of the corresponding electric potential $\varphi(x,t)$ and the electric field $\partial_x \varphi (x,t)$]{\includegraphics{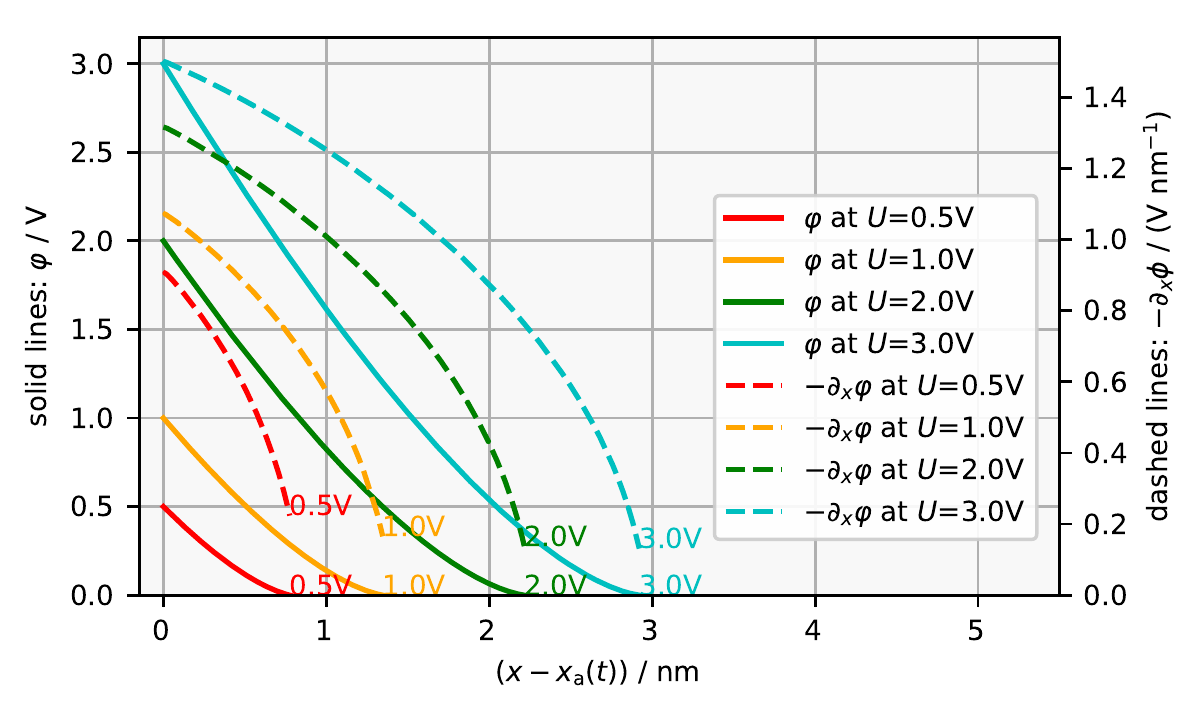}\label{fig:profiles_phi}}
		\caption{Stationary distributions of several space-dependent quantities are shown at different voltages $U$. At each voltage~$U$ there is a different stationary layer thickness~$x_\text{b}(t)-x_\text{a}(t)$ which is reflected by the respective domain of definition.}
	\end{figure}

	The main idea of the model is that the etch rate of the oxide depends on its composition.
	Thus, let us have a look at Figure~\ref{fig:profiles_n} which shows the stationary oxide composition for some voltages~$U$.
	With increasing voltage~$U$ the stationary layer thickness $x_\text{b}(t)-x_\text{a}(t)$ increases.
	Consider the four cases in Figure~\ref{fig:profiles_n}.
	The concentration of partial oxide $n_\sio(x,t)$ is the highest at the Si/SiO$_x$ interface $x_\text{a}(t) $, where the partial oxide is created.
	Closer to the SiO$_x$/solution interface~$x_\text{b}(t)$ the fraction of partial oxide becomes smaller as it is used up in Reaction~\eqref{reac:2}.
	The remaining fraction right at the SiO$_x$/solution interface~$x_\text{b}(t)$ determines the etch rate by Equation~\eqref{eq:reac2}.
	Note that this fraction of partial oxide is much smaller at the SiO$_x$/solution interface if the layer is thicker and $U$ is larger.
	This comes from an increased volume that is available for the ions~$n_\ion(x,t)$ to participate in the bulk Reaction~\eqref{reac:2}, which consumes partial oxide.
	Consequently, if $U$ is larger, the layer is etched more slowly, according to Equation~\eqref{eq:reac2}.

	The concentration of ions $n_\ion(x,t)$ is the highest at the SiO$_x$/solution interface $x_\text{b}(t)$ where the electrochemical potential of O$^{2-}$ in the oxide and the electrochemical potentials of protons and water in the electrolyte are assumed to be in equilibrium.
	With increasing distance from the SiO$_x$/solution interface~$x_\text{b}(t)$, which corresponds to going left in Figure~\ref{fig:profiles_n}, the concentration of ions decreases.
	This is because the ions enter the oxide at the SiO$_x$/solution interface~$x_\text{b}(t)$ and are then used up in the oxide volume by Reaction~\eqref{reac:2} and at the Si/SiO$_x$ interface~$x_\text{a}(t)$ by Reaction~\eqref{reac:1} and \eqref{reac:3}.
	The concentration of ions in the layer determines the curvature of the electrostatic potential $\varphi(x,t)$ in Figure~~\ref{fig:profiles_phi} by Poisson's equation. Its derivative gives us the electric field~$-\partial_x \varphi(x,t)$, which monotonically increases from about 0.2~V/nm  at the SiO$_x$/electrolyte interface to 2-3~V/nm at the Si/SiO$_x$ interface.
	
	\subsection{Dissolution valence}
\begin{figure}
		\centering
		\includegraphics{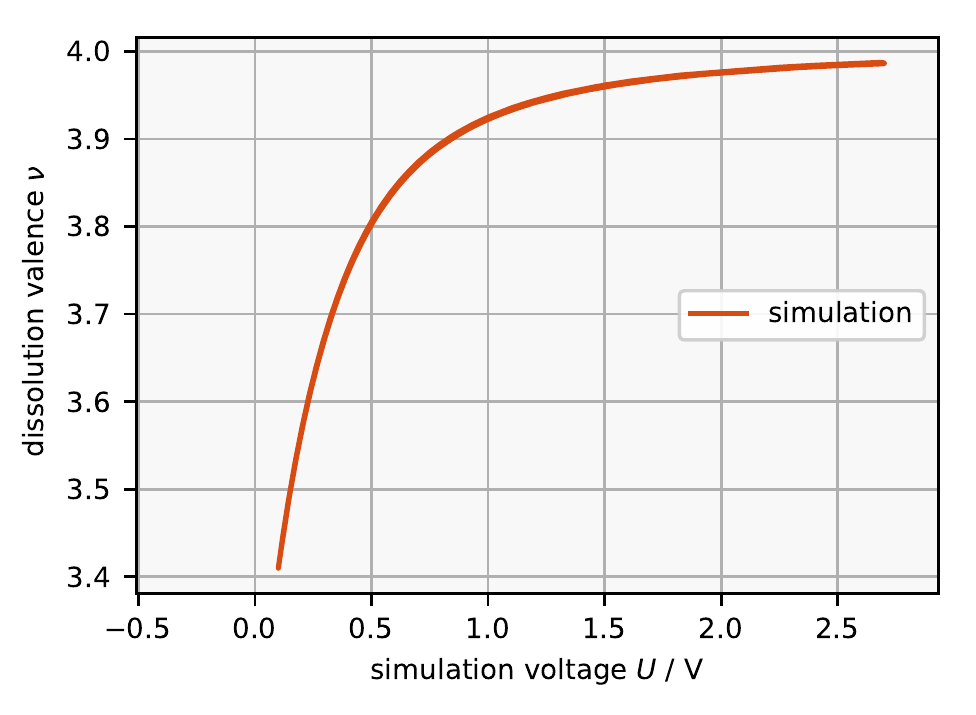}
		\caption{The dissolution valence as defined by Equation \eqref{eq:val} is plotted over the voltage $U(t)$ during simulated cyclic voltammetry using the model described in Section~\ref{sec:model} at a scan rate of 1\,mV/s, with the parameters in Table~\ref{tab:param}. As the voltage is increased the valence increases from somewhere around 3 to 4, as reported in the literature \cite{eddowes1990anodic,blackwood1992electrochemical}.}
		\label{fig:valence}
	\end{figure}

	We calculated the reaction valence $\nu(t)$ in the presented model in terms of the reaction current $I_\mathrm{reac}(t)$, the Faraday constant $F$, the Si/SiO$_x$ interface position $\xa(t)$ and the molar density of Si-atoms in the SiO$_x$ layer $\nox:$
	\begin{equation}
		\nu(t) = \frac{I_\mathrm{reac}(t)}{F\,\partial_t\xa(t)\,n_\mathrm{ox}}.
		\label{eq:val}
	\end{equation}
	The calculated reaction valence during simulated cyclic voltammetry is shown in Figure~\ref{fig:valence}.
	Due to the slow scan rate there is no noticeable hysteresis, if there is one at all.
	The simulated valence increases monotonically with the voltage from a value of about 3.4 close to the current peak to 4 at the plateau where it saturates.
	The same behavior has been confirmed by gravimetric measurements \cite{eddowes1990anodic} and by rotating ring disc measurements \cite{blackwood1992electrochemical,schonleber2012high}
	The exact value of the valence in the limit $U\rightarrow 0$ can be adjusted by changing the rate constant of Reaction~\eqref{reac:3}.
	
	
	\subsection{Impedance spectra}
	\label{sec:impedance}
	
	To obtain the impedance spectra for our model described in Section~\ref{sec:model} we set the parameter $U$ to a different function for each investigated angular frequency $\omega$, denoted by $U_\omega(t)$:
	\begin{equation}
		U_\omega(t)= 1\,\mathrm{mV} \cdot \mathrm{sin}(\omega\,t) + U_0
	\end{equation}
	The model equations were solved for many periods of $U_\omega(t)$ such that the resulting electric current $I_\omega(t)$ converged and became sine-shaped.
	We calculated the corresponding analytical signals $\widetilde{U}_\omega(t)$ and $\widetilde{I}_\omega(t)$ with the Matlab R2017b function hilbert(...) \cite{MATLAB:2017b}.
	This means that $\widetilde{U}_\omega(t)$ and $\widetilde{I}_\omega(t)$ are obtained from ${U}_\omega(t)$ and ${I}_\omega(t)$ by removing the negative and zero frequency components.
	The impedance $Z_\omega(t)$ is then calculated as
	\begin{equation}
		Z_\omega(t)=\frac{\widetilde U_\omega(t)}{\widetilde I_\omega(t)}
	\end{equation}
	which would be constant in time if the current response was perfectly linear.
	$Z_\omega(t)$ is, of course, not completely constant in our non-linear small amplitude case. Thus, we determined the value for the impedance $Z_\omega$ at a certain angular frequency $\omega$ by time-averaging $Z_\omega(t)$ over some periods.
	Alternatively one could have calculated the impedance from the values of the Fourier transformed signals at $\omega$, which is equivalent for a linear response.
	
\begin{figure}
		\subfloat[Nyquist plot of impedance spectra from a simulation using the model described in Section\,\ref{sec:model} with the parameters in table\,\ref{tab:param}.]{\includegraphics{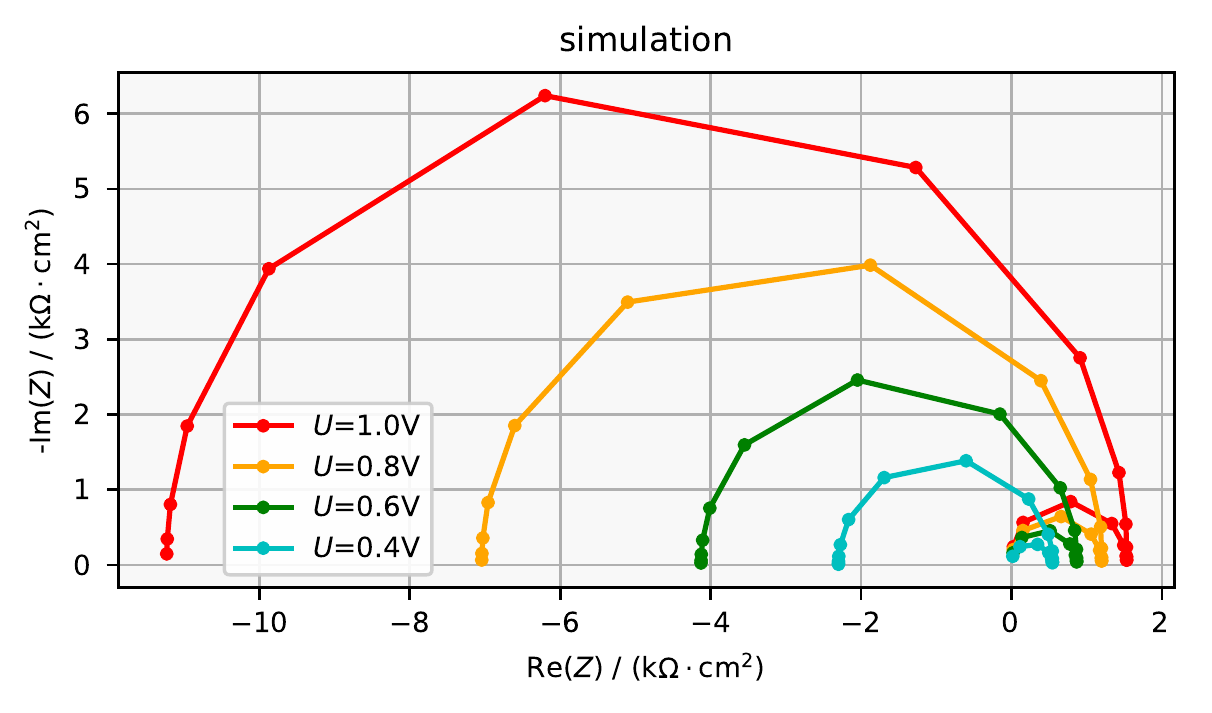}\label{fig:sim_nyq}}\\
		\subfloat[This Nyquist plot shows impedance spectra that were measured with dynamic multi frequency analysis \cite{koster2017dynamic} for (111) p-silicon of 5 to 25\,$\mathrm{\Omega}\,\text{cm}$ in a solution of 0.05M NH$_4$F and 0.025M H$_2$SO$_4$ (pH2.3).]{\includegraphics{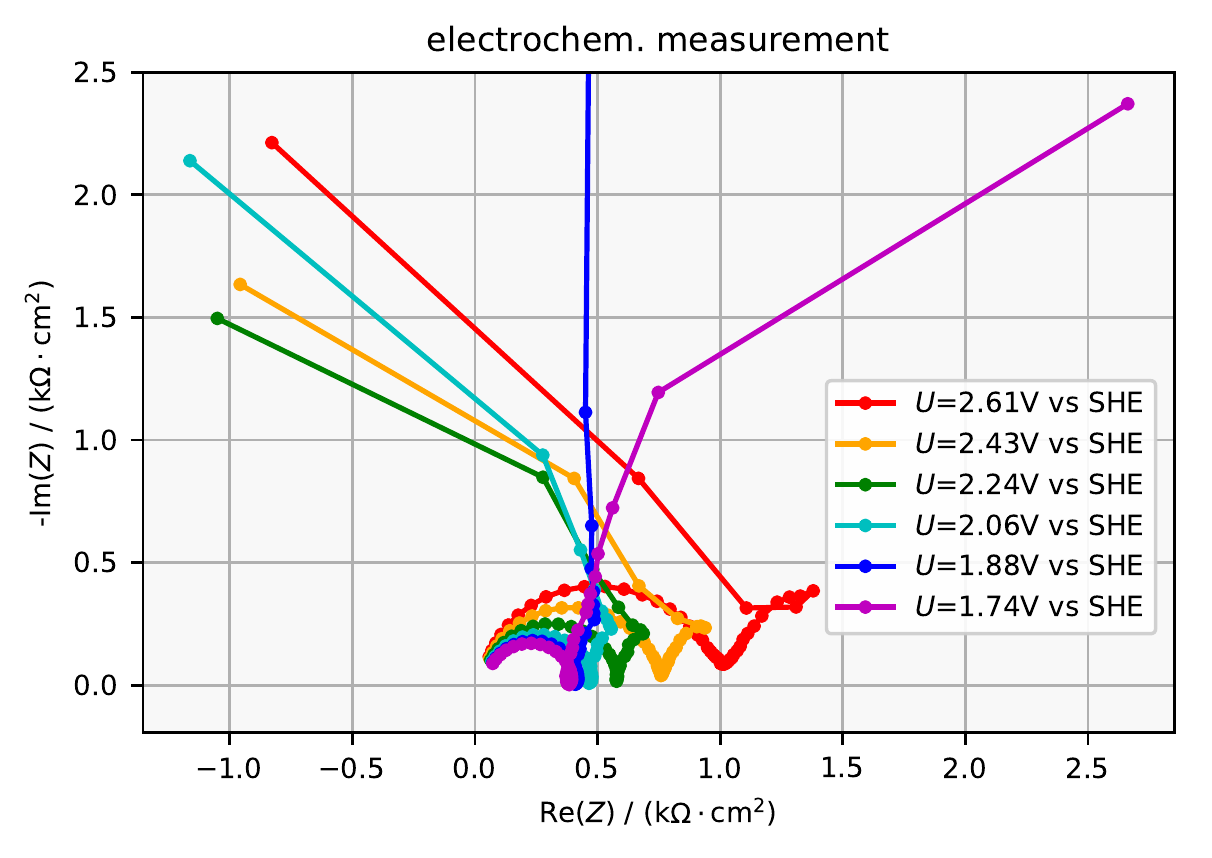}\label{fig:exp_nyq}}
		\caption{A comparison of the Nyquist plots of simulated and electrochemically measured impedance spectra. There is an offset in the voltage $U$ which is qualitatively discussed in the text (see also Fig \ref{fig:cv}). In the simulation both semi-circles grow as the voltage is increased, which is also the case in the measured spectra. The measured spectra have a `kink' which is not found in the simulated spectra. The `kink' becomes more prominent for larger voltages.}
		\label{fig:Nyq}
	\end{figure}

\begin{figure}
	\subfloat[Bode plot from the simulated data of Figure~\ref{fig:sim_nyq}.]{\includegraphics{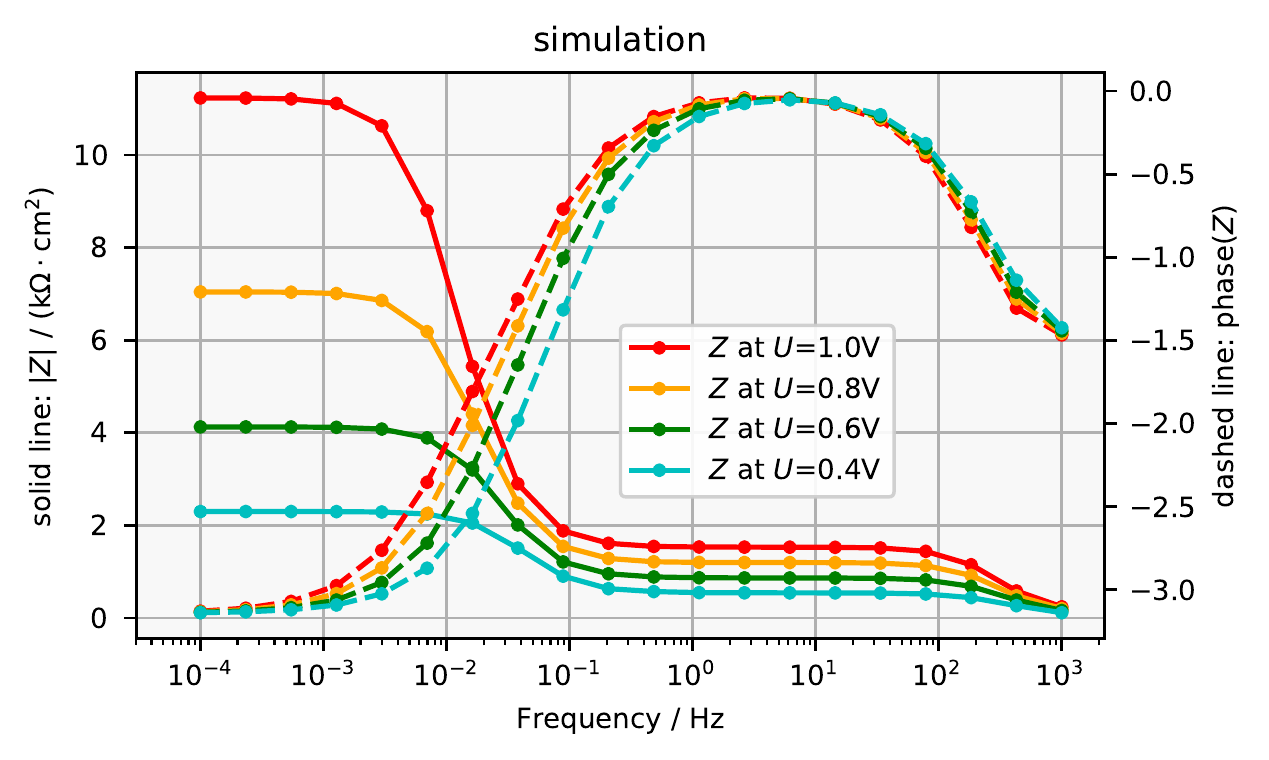}\label{fig:sim_bode}}\\
	\subfloat[Bode plot from the experimantal data of Figure~\ref{fig:exp_nyq}.]{\includegraphics{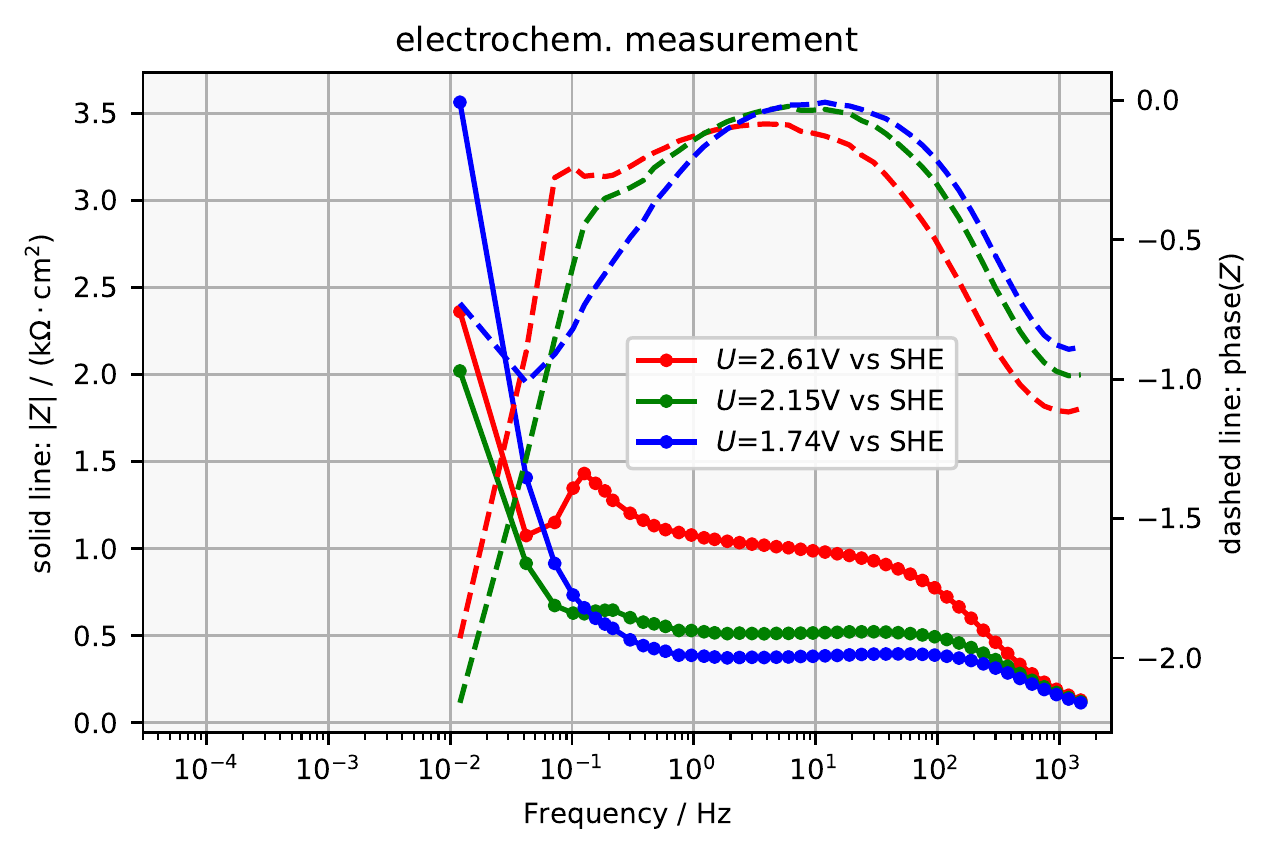}\label{fig:exp_bode}}
	\caption{Comparison of Bode plots of the same data as in Figure~\ref{fig:Nyq}.}
	\label{fig:Bode}
\end{figure}

\begin{figure}
		\centering
		\includegraphics{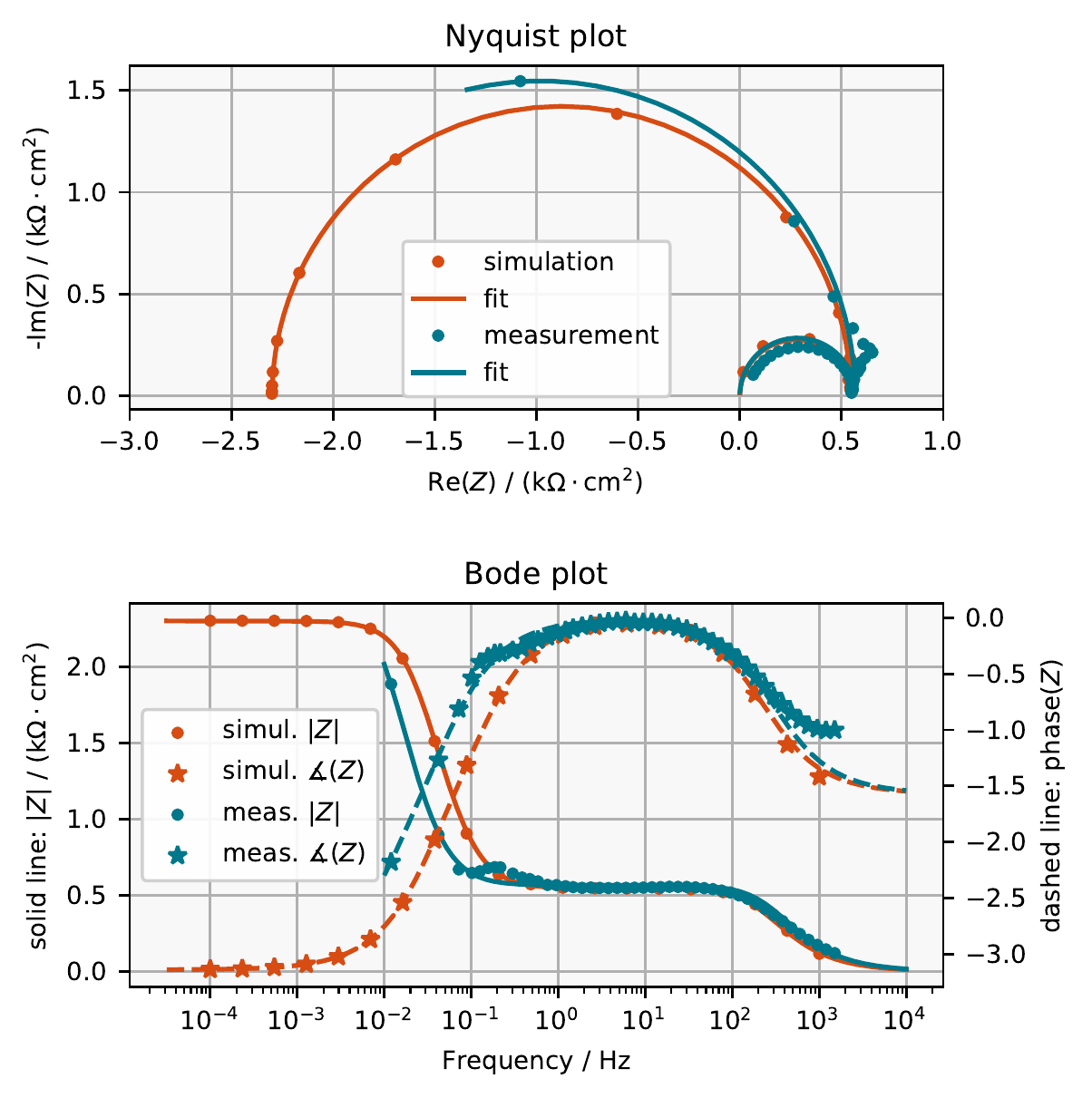}
		\caption{One simulated impedance spectrum at voltage $U=0.4$V and one measured impedance spectrum at voltage $U=2.2$V vs SHE are exemplarily compared. Fits are represented by solid lines for the absolute value $|Z|$ and by dashed lines for the complex phase of $Z$. The equivalent circuit can be seen in Figure~\ref{fig:equiv1}. The circuits fit the simulated spectrum well but do not reflect all features of the measurement. These features become more/less dominant at higher/lower voltage, as can be seen in Figure~\ref{fig:exp_nyq}}
		\label{fig:NyqSimVsExp}
	\end{figure}

	The impedance spectra obtained are plotted in Figure~\ref{fig:sim_nyq} and \ref{fig:sim_bode} and the corresponding electrochemically measured impedance spectra in Figure~\ref{fig:exp_nyq} and \ref{fig:exp_bode}. 
	The plots in Figure~\ref{fig:NyqSimVsExp} exemplarily compare one simulated impedance spectrum to one measured impedance spectrum.
	It can be seen that the simulated spectrum looks similar to the measured spectrum around 2\,V vs SHE. 

\begin{figure}
		\centering
		\includegraphics{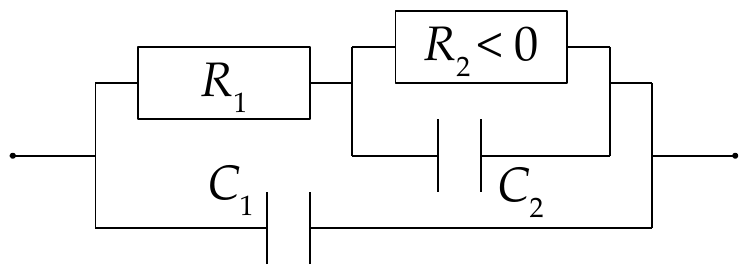}
		\caption{Equivalent circuit fitted to the spectra in Figure~\ref{fig:NyqSimVsExp}. It fits the simulated spectra quite well, but for the measured spectra it neglects the 'kink' at around 0.2Hz and it does not describe the complex phase at $\omega\rightarrow\infty$.}	   
		\label{fig:equiv1}
	\end{figure}

	The model allows us to assign a physical meaning to some features of the measured impedance spectra which are illustrated in the equivalent circuit in Figure~\ref{fig:equiv1}.
	For example, when omitting the charging term $I_\mathrm{ch}(t)$ in the numerical calculation (see Equation~\eqref{eq:current}), the small semi-circle at high frequencies in the Nyquist plot disappears.
	Thus, it is clear that the small semi-circle emerges due to interfacial charging, which means that opposing charges accumulate at the Si/SiO$_x$ interface and at the SiO$_x$/solution interface.
	Hence, any equivalent circuit would require a capacitor $C_1$, see Figure~\ref{fig:equiv1}.
	
	At an intermediate frequency at which the small and the large semi-circle merge, the impedance is almost perfectly ohmic.
	This means that this frequency is sufficiently low for the layer capacity to be negligibly small but high enough so that the oxide thickness is not affected. 
	Thus, the impedance at this frequency corresponds to the resistivity of the layer for a fixed thickness. This is represented in the equivalent circuit by a resistor $R_1$.
	Since the simulated impedance spectra at low frequencies resemble a semi-circle which ends at a real and negative value when $\omega\rightarrow 0$, we fit a capacitor $C_2\ll C_1$ in parallel with a negative resistance $R_2$. 
	This approximation seems to be valid for the electrochemically measured impedance spectra in Figure~\ref{fig:exp_nyq},  at $U\approx 2$~V vs SHE.
	
	There are some measurable features that the model does not reproduce.
	It can be seen in the Bode plot in Figure~\ref{fig:exp_bode} that at high frequencies the measured spectra approach a phase that differs from $-\pi/2$, where $-\pi/2$ would correspond to an ideal capacitor.
	This is probably caused by an uncompensated resistance in series, which leads to a complex phase of zero at $\omega\rightarrow\infty$.
	
	Another feature in the measured spectra, which is not reflected by the model, is the kink at frequencies between 0.1\,Hz and 1\,Hz. 
	This kink becomes a loop that grows for higher voltage until at some voltage it possibly causes the resonance that is known to occur in the system \cite{ozanam1992resonant}, by touching the origin in the Nyquist plot. If the impedance becomes zero at a certain frequency, it corresponds to a Hopf-bifurcation \cite{koper1994far}, which could explain the emergence of oscillations that are observed in the system \cite{turner1958electropolishing,schonleber2012high}.
	However, the loop is not explained by the model and the scenario in the electrochemical set up is probably even more complicated, as can be guessed from earlier impedance measurements at other parameters \cite{ozanam1992resonant}.
	Unfortunately, measuring the impedance of the steady state in the resonant regime appears to be challenging. 
	
	A third discrepant feature is measured at smaller anodic voltages. As the voltage is decreased in the electrochemical set up, one eventually passes the current peak and arrives at the electropolishing regime where the differential resistance is positive.
	Thus, the presumed contact point between the impedance curve and the real axis at zero frequency in the Nyquist plot moves from the left half plane to the right, as show in Figure~\ref{fig:exp_nyq}. The model, in contrast, is not valid for voltages this low, as there is no oxide layer of relevant thickness.
	
	\subsection{Origin of the negative differential resistance}
	
	From our model we can extract the necessary physical conditions leading to the negative differential resistance. First we note that in order to obtain the negative slope in the current voltage characteristics only the reactions~\eqref{eq:equilibriumreaction}, \eqref{reac:1}, \eqref{reac:2}, and \eqref{eq:dissolution} are required.
	Reaction \eqref{reac:3}, in contrast, was only introduced to adjust the valence to above 3 at very small layer thicknesses, where volume Reaction~\eqref{reac:2} hardly occurs. 
	
	Let us assume that the system is in a steady state on the negative differential branch. An increase in the voltage $U$ leads to an increase in the amount of O$^{2-}$ ions pulled into the oxide layer by reaction \eqref{eq:equilibriumreaction} which increases the rate of the first oxidation step \eqref{reac:1}, as the latter requires O$^{2-}$ ions. 
	The resulting thickening of the oxide layer works against \eqref{reac:1} because the path that the O$^{2-}$ ions have to travel in order to reach the Si/SiO$_x$ interface becomes longer. 
	Therefore, the thickening of the oxide layer partly compensates the rate increasing effect of the larger voltage $U$ on the first oxidation step~\eqref{reac:1}. 
	The the second oxidation step \eqref{reac:2}, however, which happens inside the layer volume, is supported by the thicker layer, or at least it is less suppressed compared to the first oxidation step \eqref{reac:1}, which happens at the Si/SiO$_x$ interface.
	Thus, the oxide layer now contains a higher fraction of fully oxidized silicon, SiO$_2$, as compared to a smaller applied voltage~$U$. 
	Consequently, the oxide layer is etched more slowly by Reaction~\eqref{eq:dissolution}.
	This thickens the oxide layer even further such that the rate increasing effect of the larger voltage $U$ on the first step \eqref{reac:1} is overcompensated. 
	
	Note that without considering reaction \eqref{reac:2}, any increase in voltage is compensated by a thicker oxide layer in such a way that the electric field inside the oxide layer as well as the current remain constant, independently of the ratios or the absolute vales of the rate constants of SiO \eqref{reac:1} and SiO$_2$ \eqref{reac:3} formation.
	
	In short, the negative differential resistance is the result of the interplay between the further oxidation of partially oxidized SiO$_x$ inside the oxide layer and the dependence of the etch rate on the stoichiometry of the oxide.

	\section{Conclusion}
	\label{sec:conclusion}
	We presented a physical model that reproduces the cyclic voltammogram in the region of negative differential resistance for voltages below the resonant current plateau, the monotone increase in the valence, and the monotone increase in the oxide layer thickness.
	Moreover, our simulations reproduce our measured impedance spectra, which were obtained via dynamic multi-frequency analysis. 
	
	The model is build on principles of non-equilibrium thermodynamics, solid state chemistry and semi-conductors physics.
	It is constructed in a way to capture the behavior qualitatively and semi-quantitatively with minimal complexity. To this end, we could reproduce all measured trends assuming that the reaction is completely limited by migration of ions in the oxide layer. Hence, potential dependences of the electrochemical reaction rates or changes in the potential drops across the space charge layer in the Si or the Helmholtz layer could be neglected. On the other hand, a dependence of the chemical etch rate of SiO$_x$ on the oxide stoichiometry and on the (slow) further oxidation of SiO$_x$ within the oxide layer are crucial elements for reproducing the negative differential branch in the current-voltage characteristic.
	
	This model presents a basis on which one can build to further uncover nonlinear features of the Si electrodissolution dynamics, such as current oscillations or the patterns observed with n-type Si electrodes. For this it is likely that some of the simplifying assumptions entering our base model have to be relaxed.
	
	\bibliographystyle{epj} 
	\bibliography{literatur}
	
	
\end{document}